\documentclass[prl,aps,twocolumn,amsmath,amssymb]{revtex4}

\def\gsim{ \lower .75ex \hbox{$\sim$} \llap{\raise .27ex \hbox{$>$}} }

\def\lsim{ \lower .75ex \hbox{$\sim$} \llap{\raise .27ex \hbox{$<$}} }

\def\IZ{\relax\ifmmode\mathchoice
{\hbox{\cmss Z\kern-.4em Z}}{\hbox{\cmss Z\kern-.4em Z}}
{\lower.9pt\hbox{\cmsss Z\kern-.4em Z}} {\lower1.2pt\hbox{\cmsss
Z\kern-.4em Z}}\else{\cmss Z\kern-.4em Z}\fi}
\def\IR{\relax{\rm I\kern-.18em R}}

\def\one{{\hbox{ 1\kern-.8mm l}}}

\newlength{\bredde}
\def\slash#1{\settowidth{\bredde}{$#1$}\ifmmode\,\raisebox{.15ex}{/}
\hspace*{-\bredde} #1\else$\,\raisebox{.15ex}{/}\hspace*{-\bredde}
#1$\fi}
\newcommand{\ft}[2]{{\textstyle\frac{#1}{#2}}}

\newsavebox{\zzzbar}
\sbox{\zzzbar}
  {\setlength{\unitlength}{0.9em}
  \begin{picture}(0.6,0.7)
  \thinlines
  \put(0,0){\line(1,0){0.6}}
  \put(0,0.75){\line(1,0){0.575}}
  \multiput(0,0)(0.0125,0.025){30}{\rule{0.3pt}{0.3pt}}
  \multiput(0.2,0)(0.0125,0.025){30}{\rule{0.3pt}{0.3pt}}
  \put(0,0.75){\line(0,-1){0.15}}
  \put(0.015,0.75){\line(0,-1){0.1}}
  \put(0.03,0.75){\line(0,-1){0.075}}
  \put(0.045,0.75){\line(0,-1){0.05}}
  \put(0.05,0.75){\line(0,-1){0.025}}
  \put(0.6,0){\line(0,1){0.15}}
  \put(0.585,0){\line(0,1){0.1}}
  \put(0.57,0){\line(0,1){0.075}}
  \put(0.555,0){\line(0,1){0.05}}
  \put(0.55,0){\line(0,1){0.025}}
  \end{picture}}

\def\Im{{\rm Im ~}}

\newcommand{\ena}{\end{eqnarray}}
\newcommand{\beqa}{\begin{eqnarray}}
\newcommand{\eeqa}{\end{eqnarray}}
\newcommand{\bea}{\begin{eqnarray}}
\newcommand{\eea}{\end{eqnarray}}

\newcommand{\be}{\begin{equation}}
\newcommand{\ee}{\end{equation}}
\def\ft#1#2{{\textstyle{\frac{\scriptstyle #1}{\scriptstyle #2} } }}
\def\fft#1#2{{\frac{#1}{#2}}}

\usepackage{graphicx}

\def\im{{{\rm i}}}
\def\nn{\nonumber}
\def\ben{\begin{equation}}
\def\een{\end{equation}}

\def\bea{\begin{eqnarray}}
\def\eea{\end{eqnarray}}
\def\be{\begin{equation}}
\def\ee{\end{equation}}
\def\beq{\begin{eqnarray}}
\def\eeq{\end{eqnarray}}

\def\({\left (}
\def\){\right )}
\def\[{\left [}
\def\[{\right ]}

\def\ba{\begin{eqnarray}}
\def\ea{\end{eqnarray}}

\def\del{{\partial}}

\input amssym.def
\input amssym.tex

\usepackage{dcolumn}

\begin{document}

\begin{flushright}
UPR-1301-T\ \ \ MI-TH-1943
\end{flushright}

\title{Positive Energy Functional for Massless Scalars in Rotating 
Black Hole Backgrounds of Maximal Ungauged Supergravity }

\author{M. Cveti\v c$^{1,2}$,  G.W. Gibbons$^3$, C.N. Pope$^{4,3}$ 
 and B.F. Whiting$^5$}

\affiliation{\vskip 0.3cm
{\it $^1$ Department of Physics and Astronomy,
University of Pennsylvania, Philadelphia, PA 19104, USA}
\vskip .5mm
{\it $^2$ Center for Applied Mathematics and Theoretical Physics, University of Maribor, Maribor, Slovenia}
\vskip .5mm
{\it $^3$ DAMTP, Centre for Mathematical Sciences, Cambridge University, Wilberforce Road, Cambridge, CB3 0WA, UK}
\vskip 0.5mm
{\it $^4$ George P. \& Cynthia W. Mitchell Institute for Fundamental Physics and Astronomy, Texas A\&M University, College Station, TX 77843-4242, USA}
\vskip .5mm
{\it $^5$ Department of Physics, P.O. Box 118440, University of Florida,
Gainsville, FL 32611-8440}
\vskip 0.5mm}

\begin{abstract}
 
We outline a proof of
the stability of a massless neutral scalar
field $\psi$ in the background of  a wide class of four dimensional
asymptotically flat rotating and ``electrically
charged'' solutions of supergravity, and the low energy limit of string
theory, known as STU metrics.
Despite their complexity, we find it possible to circumvent 
the difficulties presented by the existence
of ergo-regions and the related phenomenon of super-radiance
in the original metrics by following 
a strategy due to Whiting,
and passing to  an auxiliary
metric admitting an everywhere lightlike Killing field
and constructing a scalar field $\Psi$ (related to a possible unstable mode $\psi$
by a non-local transformation) which satisfies  the
massless wave equation with respect
to the auxiliary metric.  By contrast with the case for $\psi$, 
the associated energy density of $\Psi$ is not only conserved
but is also non-negative.

\end{abstract}


\maketitle


Recent successes of the LIGO and Virgo gravitational wave detectors 
\cite{Abbott:2016blz}, and of the Event Horizon Telescope \cite{eht}, have 
triggered renewed interest in probing the near horizon geometry of black 
holes, and in testing the predictions of General Relativity against alternative 
theories of gravity.  It is important that such alternative theories i) have a 
well-posed initial value formulation, ii) that proposed matter fields have 
an energy momentum tensor satisfying sensible energy conditions and, 
ideally, iii) for any black hole solutions obtained, that some minimum 
requirement of stability holds.  An example obtained from String Theory 
and that meets the first two of these requirements is the Kerr-Sen metric 
\cite{sen}, which generalized the Kerr-Newman spacetime, and has recently 
been the subject of much investigation in relation the recent results from 
observations with the event horizon telescope \cite{eht}.  
The supergravity-inspired STU metrics we study further generalize 
the Kerr-Sen metric, and the stability result we obtain applies equally well to it.  
STU theories are particularly interesting because they are Lagrangian based \cite{chcvlupo} 
and their matter sector separately satisfies the weak, strong and dominant energy 
conditions (see \cite{Hawking:1973uf} for definitions) in common with all ungauged 
supergravity theories \cite{Gibbons:2001gy}.  Thus, they have the potential to provide 
a genuine, physically-testable, mathematically consistent, viable alternative against 
which to evaluate the predictions of General Relativity as a theory of gravity.

The stability of black hole spacetimes within General Relativity has been of 
longstanding interest since the work of Regge and Wheeler \cite{Regge:1957td} 
in 1957, and Zerilli \cite{Zerilli:1970se} and Vishveshwara \cite{Vishveshwara:1970cc} 
in 1970, on the Schwarzschild spacetime, and the initial work of Press and 
Teukolsky \cite{Teukolsky:1973ha,Press:1973zz} for the Kerr black hole in 1973.  
These papers were all concerned 
with mode stability: whether small perturbations can grow exponentially.
A deeper analysis is required to establish whether linear perturbations 
decay generically at late times \cite{Dafermos:2016uzj}, and whether linear 
perturbations can excite non-linear growth \cite{Klainerman:2017nrb}.  A
subtlety that arises for black holes is that even if
the solution settles down to a stationary final state, it may not have
the same mass and angular momentum as the solution around which 
one is linearising.  The final solution should, however, by the
no-hair theorems, be a regular stationary black hole within the space of 
solutions of the theory under consideration.

  When one considers the stability of the rotating Kerr solution \cite{kerr}, one
encounters the difficulty of dealing with super-radiance.  
Two key steps in understanding the stability
of the the Kerr solution were the separation of the relevant 
perturbation equations
\cite{Teukolsky:1972my} 
and the construction, using properties of the resultant confluent Heun equations,
of a modified wave equation admitting a conserved positive
energy \cite{Whiting:1988vc}. The latter has played an important role
in a recent advances in this subject \cite{Dafermos:2014cua,daCosta:2019muf}.

The extension to the rotating electrically charged Kerr-Newman solution
has been impeded by the lack of separation results analogous to those of
\cite{Teukolsky:1972my}. 
For the
special case of an electrically neutral massless scalar wave equation 
on a fixed Kerr-Newman background, see 
\cite{Civin:2014bha,Civin2}.  Very recently,  the Teukolsky equations 
for  Kerr-Newman spacetime have been obtained \cite{Giorgi:2020ujd} 
as a first, very preliminary, step towards being able to discuss the 
stability question in this difficult case.

The main 
purpose of the present letter is to report on the mode stability of
the massless wave equation on the fixed background of 
a class of four-dimensional four-charged rotating black hole 
solutions \cite{cvetyoum,chcvlupo} of ungauged
supergravity theory known as ``STU black holes,'' which are 
generalisations of the Kerr-Newman black hole, 
admitting a separation of variables \cite{Cvetic:1997xv}.  STU
supergravity is a consistent truncation of maximal four-dimensional
supergravity 
having four abelian gauge fields, and the four ``charges''
of the black holes we consider here are carried by these fields.
If the charges are set equal, the solution reduces to the Kerr-Newman 
black hole \cite{Chow:2014cca}.
The four-dimensional supergravity can be embedded within the 
low-energy limit of string theory, by means of a dimensional reduction on
a 6-torus.
For an account of the existence of Killing Staeckel and Killing-Yano tensors
extending known results in the Kerr case, 
see \cite{Keeler:2012mq}, \cite{Chow:2014cca} and \cite{Frolov}.  

In previous work \cite{Cvetic:2016bxi,Cvetic:2017zde}, 
three of us have advocated
the use of these metrics as a foil  to assess the accuracy with
which astronomical  observations of black holes can confirm
that the metric is given by that of Kerr \cite{kerr}. 
Quite apart from the STU black holes being solutions within supergravity or
string theory, they have the merit that one can view them as a parameterised 
family of generalisations of the Kerr black hole for which, ungauged, their 
accompanying mater sector obeys the dominant,  
strong and weak energy conditions of general relativity.  
A further distinct advantage they have over many rival versions of modified 
gravity is that STU metrics can be obtained from a Lagrangian and so have 
a Hamiltonian and an initial-value formulation.  
To be effective as a foil, it is vital to understand whether STU 
black holes are stable, failing which they could not become the endpoints 
of gravitational wave emission arising from the inspiral and merger 
of massive black holes, as have recently been observed by the LIGO and 
Virgo collaborations, starting with GW150914 in 2015 
\cite{Abbott:2016blz}.  
A similar reliance on stability applies
for the use of STU black holes (in particular the special case of the
Sen \cite{sen} black hole) in relation to observations made by the event horizon telescope 
\cite{eht}.

An outgoing mode solution is one that has no support on the past horizon
or at past null infinity. With the separation of variables given in
(\ref{mode_decomp}), an outgoing mode with complex $\omega$ would grow
exponentially in time if~$\Im{\omega}>0$, and would therefore be
considered ``unstable.'' Outgoing modes where $\omega$ is real and not
equal to 0, if they exist, would
also be considered unstable, since they would present an obstacle to
proving decay at late times ($\omega=0$ is discussed further below). 
Outgoing modes with $\Im{\omega}<0$ decay
exponentially in time, and are ``quasi-normal modes.''

  We shall follow the strategy employed for the Kerr metric 
in \cite{Whiting:1988vc}.  We start with 
the separated solutions of the massless wave equation $\Box_g \psi =0$,
where $\Box_g$ is the covariant D'Alembertian of the stationary STU 
black hole metric $g$, and then construct a new function $\Psi$ satisfying
$\Box_{\hat g} \Psi =0$,
where $\Box_{\hat g}$ is the covariant D'Alembertian of an auxiliary
metric $\hat g$ with a null Killing vector field $K^\mu$.  

By  contrast with $\psi$ and the physical metric
$g_{\mu \nu}$,  the conserved energy current $J^\mu=   -K^\nu\, 
T^\mu{}_\nu  \label{Jdef}$
has a positive  energy density, $J^0 \ge 0$,
where $T_{\mu \nu} =\del_\mu\Psi\, \del_\nu\Psi -
  \ft12 \hat g_{\mu\nu}\, \hat g^{\rho\sigma}\,
  \del_\rho\Psi\, \del_\sigma\Psi$ is the energy-momentum tensor
with respect to the auxiliary metric $\hat g_{\mu \nu}$.  In 
\cite{Whiting:1988vc} it was shown  that  establishing positivity of  an
energy functional,
albeit with respect to the auxiliary metric $\hat g_{\mu\nu}$, of the
auxiliary field $\Psi$ suffices to show mode stability
of the original massless scalar field $\psi$. The relationship between
$\Psi$ and $\psi$ is non-local, being effected by means of a
pair of certain integral and differential transforms, 
chosen so that an unstable mode for $\psi$ maps to an unstable mode for $\Psi$.

   For the auxiliary metric $d\hat s^2=\hat g_{\mu\nu}dx^\mu dx^\nu$,  
we have 
\bea
d\hat s^2 &=& \Omega^2\,\Big( \fft{dr^2}{\Delta} + d\theta^2 \Big)+ 
\fft{(P + a^2\cos^2\theta)\,\Delta\sin^2\theta}{
    a^2\,\Omega^2} \, d\phi^2\nn\\
&& 
 -\fft{2\,\Delta^{1/2}\, \sin\theta}{a}\, dt d\phi\,,\label{auxmet}
\eea
where $\Delta$ is given in (\ref{metfuns}), and
$P$ is given in (\ref{Pdef}), and $\Omega$ is given in 
(\ref{Omega}).  The function $P$ is manifestly positive
for all $r\ge r_+$ (the radius of the outer horizon). This implies that
with respect to the auxiliary metric $\hat g_{\mu \nu}$
the azimuthal Killing vector $\del_\phi$ is everywhere spacelike, 
and the Killing vector $\del_t$ is
everywhere lightlike. By contrast with respect to the physical
STU metric $g_{\mu \nu}$  the Killing vector $\del_t$ is timelike
outside the ergosphere upon which it becomes null and spacelike
within the ergosphere.

The fact that   $\del_t$ is lightlike with respect to $\hat g_{\mu \nu}$
and the fact
that $T_{\mu\nu}$ satisfies
the dominant energy condition, implies the positivity 
of the energy density $J^0$.  Specifically, from $J^\mu=-K^\nu\, T^\mu{}_\nu$ 
and $K=\del_t$,
the total energy $\int \sqrt{-\hat g}\,d^3x \,J^0$ is
\bea
\fft1{2a}\, \int \sin\theta\, dr d\theta d\phi\,&& \!\!\!\!\!\Big[
  (P + a^2\cos^2\theta)\, (\del_t\Psi)^2 +
 \Delta\, (\del_r\Psi)^2\nn\\
&& + (\del_\theta\Psi)^2\Big]\,,
\eea
which is manifestly positive, reducing to that for the Kerr
metric in \cite{Whiting:1988vc} if the ``electric charges'' are set to zero.  

  We shall now present a brief summary of the key steps in the derivation
of the auxiliary metric (\ref{auxmet}).  Further details will be given in
a longer paper to follow \cite{longpaper}.

    The four-charge rotating STU black hole solution was obtained in
\cite{cvetyoum}; a convenient form for the metric is \cite{chcvlupo}:
\bea\label{STU-metric}
ds^2&=&-{\rho^2-2\mu r\over W}\left(dt+ \bar\omega d\phi\right)^2\nn\\
&&+
  W\left({dr^2\over \Delta}+d\theta^2+
{\Delta\sin^2\theta d\phi^2\over \rho^2-2\mu r}\right)
\eea
where 
\bea
W^2&=&r_1r_2r_3r_4+a^4\, \cos^4\theta + 
\Big[2r^2+2\mu r\Sigma_i^2+8\mu r\Pi_c\Pi_s\nn\\
&&\quad\qquad\qquad\qquad -
 4\mu^2\left(\Sigma_{ijk}^2+2\Pi_s^2\right)\Big]\,a^2\,\cos^2\theta\,, \nn\\
r_i&=&r+2\mu s_i^2,\quad \rho^2=r^2+a^2\cos^2\theta,\nn\\
\Delta&=&r^2-2\mu r + a^2
  =(r-r_-)(r-r_+)\,,\nn\\
\bar\omega &=& {2\mu a\,[r\Pi_{c}-(r-2\mu)\Pi_{s}] \sin^2\theta
                       \over \rho^2-2\mu r}
\,,\nn\\
c_i&=&\cosh \delta_{i}\,, \quad s_{i}=\sinh \delta_{i}\,,\nn\\
\Pi_c&=& c_1 c_2 c_3 c_4\,,\quad \Pi_s=s_1 s_2 s_3 s_4\,,\nn\\
\Sigma_i^2&=&s_1^2+s_2^2+s_3^2+s_4^2\,,\quad
\Sigma_{ijk}^2=\sum_{i<j<k} s_i^2s^2_js_k^2\,. \label{metfuns}
\eea
   
Changing variables from $r$ and $\theta$ to
\be
x=\fft{r-r_-}{r_+-r_-}\,,\qquad y= \ft12 (\cos\theta+1)\,,\label{xydef}
\ee
the massless scalar wave equation $\square\psi=0$ can be separated by writing
\be
\psi = e^{-\im \omega t}\, e^{\im m\phi}\, \fft{X(x)}{\sqrt{x(x-1)}}\,
  \fft{Y(y)}{\sqrt{y(y-1)}}\,,\label{mode_decomp}
\ee
with $X(x)$ and $Y(y)$ satisfying
\be
\fft{X''(x)}{X(x)} + V_x =0\,,\qquad 
\fft{Y''(y)}{Y(y)}+ V_y =0\,,\label{XYeqns}
\ee
where a prime denotes a derivative with respect to the argument, and
\bea
V_x&=& -\tilde\alpha^2 + 
\fft{\tilde\alpha\tilde\kappa +\tilde\lambda+\ft12\tilde\kappa^2}{x} +
\fft{\ft14-\tilde\beta^2}{x^2}\nn\\
&&
+\fft{\tilde\alpha\tilde\kappa -\tilde\lambda-\ft12\tilde\kappa^2}{x-1} +
\fft{\ft14-\tilde\gamma^2}{(x-1)^2}\,,
\eea
\bea
V_y&=&-\alpha^2 +
\fft{\alpha\kappa +\lambda+\ft12\kappa^2}{y} +
\fft{\ft14-\beta^2}{y^2} \nn\\
&&
+\fft{\alpha\kappa -\lambda-\ft12\kappa^2}{y-1} +
\fft{\ft14-\gamma^2}{(y-1)^2}\,.\label{VxVy}
\eea
The constants in the potential $V_x$ are given by
\bea
\tilde\alpha&=& \im \omega\, (r_+ - r_-)\,,\quad
\tilde\kappa=-\fft{\im}{4}\, \omega\, (r_++r_-)\, \sum_i \cosh2\delta_i\,,\nn\\
\tilde\beta &=& \fft{\im\, [a m - \omega(r_+ + r_-)
                      (\Pi_c\, r_- + \Pi_s\, r_+)]}{
                     r_+ - r_-}\,,\nn\\
\tilde\gamma&=&-\fft{\im\, [a m - \omega(r_+ + r_-)
              (\Pi_c \, r_+ + \Pi_s \, r_-)]}{
                     r_+ - r_-}\,,\nn\\
\tilde\lambda &=& \fft12 + \tilde\alpha (\tilde\gamma-\tilde\beta) -
  \fft12 (\tilde\gamma -\tilde\beta)^2 +\lambda_T + \nu\,,\label{tilde_params}
\eea
(from which $\omega_+=a/[(r_++r_-)(\Pi_c \, r_+ + \Pi_s \, r_-)]$) with
\bea
\nu&=&\fft{\omega^2\,(r_++ r_-)^2}{32}\, \Big[
 \sum_i\tilde c_i^2 - 6\sum_{i<j} \tilde c_i\, \tilde c_j\nn\\
&& +
   16 (\Pi_c-\Pi_s-1)^2 + 32(2\Pi_c-1)\Big]\label{nudef}\,,
\eea
where $\tilde c_i=\cosh2\delta_i$;  and in $V_y$ we have
\bea
\alpha&=&2 a\omega\,,\qquad \kappa=0\,,\qquad \beta= - \fft{m}{2}\,,\qquad
\gamma= \fft{m}{2}\,,\nn\\
\lambda&=& \fft12 + \alpha (\gamma-\beta) - \fft12 (\gamma-\beta)^2
  + \lambda_T\,.\label{whitingeq0}
\eea
In these expressions, $\lambda_T$ is the analogue of the
separation constant employed in \cite{Teukolsky:1972my} 
in the case of the Kerr metric.  Note that  
$\omega=m\omega_+$ is the super-radiant threshold, where $\tilde\gamma$
in (\ref{tilde_params}) changes sign.

An unstable mode would have:
\be
X(x)\sim
\begin{cases}
e^{\tilde{\alpha}x}x^{-\tilde{\kappa}}& {\rm as}~x\rightarrow\infty,\\
(x-1)^{-\tilde{\gamma}}& {\rm as}~x\rightarrow 1.
\end{cases}
\ee
  Following the strategy in 
  \cite{Whiting:1988vc}, we 
define a new radial function $\tilde h(x)$ by means of the integral
transform ($\Im{\omega}>0$)
\bea
\tilde h(x) &=& e^{\hat \alpha x}\,
x^{\hat\beta}\, (x-1)^{\hat\gamma}\times\nn\\
&&\!\!\!\!\!\!\!\!\!\!\!\!\!\!\int_1^\infty \!dz
\, e^{2\tilde\alpha x z}\, e^{-\tilde\alpha z}\,z^{-\ft12-\tilde\beta}\,
 (z-1)^{-\ft12-\tilde\gamma}\, X(z)\,,
\eea
where
\bea
\hat\alpha&=&-\tilde\alpha\,,\quad \hat\kappa=-\tilde \beta +\tilde\gamma \,,
\quad \hat\beta=-\ft12(\tilde\beta+\tilde\gamma+\tilde\kappa)\,,\nn\\
\hat\gamma&=&-\ft12(\tilde\beta+\tilde\gamma-\tilde\kappa)\,,\quad
\hat\lambda=\tilde\lambda\,.\label{hatted}
\eea
The function $\widetilde X(x)=\sqrt{x(x-1)}\, \tilde h(x)$ satisfies
the same equation as does $X(x)$ in (\ref{XYeqns}), except that
$V_x$ in (\ref{VxVy}) is now given by replacing the tilded constants
by the hatted ones given in (\ref{hatted}) \cite{Whiting:1988vc}, 
and has been chosen so that an unstable mode for $X(x)$ maps to an unstable mode for $\widetilde{X}(x)$.
  With these transformations, the radial equation becomes
\bea
&&x(x-1) \tilde{h}'' + (2x-1) \tilde{h}' \nn\\
&&~~~~\,+
  [\omega^2\, P(x)+ \omega^2\, a^2
 - 4a m \omega\,x -\lambda_T]\, \tilde{h}=0 \,,~~~~~
\label{heqn}
\eea
where
\bea
P(x) &=& (r_+ + r_-)^2\, \Big[
\fft{\Big(\sum_i \tilde c_i -4 \Pi_c + 4 \Pi_s\Big)^2\,(x-1)}{64x}
 \nn\\
&&+
 \fft{\Big(\sum_i \tilde c_i +4 \Pi_c - 4 \Pi_s\Big)^2\,x}{64 (x-1)}
  + 2(\Pi_c+\Pi_s)x \Big]\nn\\
&& 
 + (r_+ - r_-)^2\, x(x-1)-a^2 \,,\label{Pdef}
\eea
and the function $P(x)$ is manifestly positive 
for all $x\ge1$.  

  In a similar vein, one may introduce an angular function $v(y)$, this time
related
to $Y(y)$ via a differential transform, with $\widetilde Y(y)=\sqrt{y(y-1)}\, 
v(y)$ satisfying the same equation as does $Y(y)$ in (\ref{XYeqns}), except
that $V_y$ in (\ref{VxVy}) is now given by replacing the constants 
$\alpha$, $\beta$, $\gamma$, $\kappa$ by barred ones, given by
\cite{Whiting:1988vc} (Note:  $\lambda$  is unaffected)
\be
\bar\alpha =\alpha\,,\quad \bar\kappa=-\beta+\gamma\,,\quad
\bar\beta=\ft12 n + \beta \,,\quad\bar\gamma= \ft12 n -\gamma\,,
\ee
where the integer $n$ is equal to $|m|$. Taking $m$, without loss of 
generality, to be non-negative, $v$ satisfies
\bea
&&y(y-1) v'' + (2y-1) v' \label{veqn}\\
&&+ [4 a^2\, \omega^2\, y(1-y) + 4 a m \omega\,
(y-1) - \lambda_T]\, v=0\,.\nn
\eea

  We now perform an ``unseparation of variables,'' defining a
wave function $\Psi(t,x,y,\phi)=e^{-\im \omega t + \im m \phi}\,
   \tilde h(x)\, v(y)$.  
Expressed
in terms of the original radial and angular variables
$r$ and $\theta$ by using (\ref{xydef}), $\Psi$ satisfies
the wave equation
\bea
&&\Big[\del_r(\Delta\del_r)+
 \fft1{\sin\theta}\,\del_\theta(\sin\theta\,\del_\theta)
-(P + a^2\cos^2\theta) \,\del_t^2\nn\\
&&\ \  -2a\Big(\cos\theta +
\fft{r-\mu}{\epsilon_0\, \mu}\Big) \,\del_t\, \del_\phi\Big]\, \Psi=0\,,
\label{neweq}
\eea
where $\epsilon_0= (r_+ - r_-)/(r_+ + r_-)$.  (Note:  
$\lambda_T$ cancels between the radial and
angular terms in the unseparation process.)  This generalises
the result given in \cite{Whiting:1988vc} for the Kerr metric, 
the difference being the replacement of the function $f$ in that paper
by the function $P$ in (\ref{neweq}).  
By pursuing techniques similar to those used in 
\cite{Shlapentokh-Rothman:2013hza,Andersson:2016epf,daCosta:2019muf}, 
our results for $\Im{\omega}>0$ extend to real $\omega\ne 0$.  For $\omega = 0$, $\psi$ in (\ref{mode_decomp}) becomes time independent. The
radial equation then reverts to the hypergeometric equation, with three regular
singular points $r_-$ , $r_+$ and $\infty$, and requires a separate 
discussion. In the Kerr
limit, Teukolsky \cite{Teukolsky:1972my} has shown that
this equation does not have solutions which are well-behaved both at the
horizon and at infinity (see also \cite{Press:1973zz}).  
By a similar method, that conclusion also holds in
the present case. 

  One can read off, up to an arbitrary conformal scaling, an inverse metric
$\hat g^{\mu\nu}$ such that the terms involving two derivatives of $\Psi$ in
(\ref{neweq}) are of the form $\hat g^{\mu\nu}\, \del_\mu\, \del_\nu\, \Psi$.
Thus, the non-zero components are of the form
\bea
\hat g^{rr} &=& \Delta\, \Omega^{-2}\,,\qquad
\hat g^{\theta\theta}= \Omega^{-2}\,,\nn\\
\hat g^{tt}&=& -(P + a^2\, \cos^2\theta)\, \Omega^{-2}\,,
\nn\\
\hat g^{t\phi}&=& -a \Big(\cos\theta+ \fft{r-\mu}{\sqrt{\mu^2-a^2}}\Big)\,
\Omega^{-2}\,.
\eea
One may then look for a choice of the conformal factor such that the
remaining terms in the wave equation (\ref{neweq}) are reproduced also;
that is, that a multiple of (\ref{neweq}) may be written as the covariant D'Alembertian 
\be
\fft1{\sqrt{-\hat g}}\, \del_\mu\Big(\sqrt{-\hat g}\, \hat g^{\mu\nu}\,
\del_\nu \Psi\Big)=0\,.
\ee
That an $\Omega$ exists is non-trivial, since the
same function $\Omega$ must produce both the $(\del_r\Delta)\,\del_r\Psi$ term
and the $\cot\theta\, \del_\theta\Psi$ term.  
In fact, an $\Omega$ does exist, and can be 
given by
\be
\Omega^2= \Delta^{1/2}\, \Big(\cos\theta+ \fft{r-\mu}{\sqrt{\mu^2-a^2}}\Big)
\, \sin\theta\,.\label{Omega}
\ee
The 
auxiliary metric itself is given in (\ref{auxmet}).  
An extended version of our work presented here is in preparation \cite{longpaper}.

To appreciate more fully the complexity of the STU metrics, it is necessary to 
expand completely out the compact expressions presented in (\ref{metfuns}) 
and (\ref{Pdef}), and used in (\ref{STU-metric}) and (\ref{neweq}).  It should be 
recognized that any proof of stability for STU metrics may require a process 
at least as complicated as that which has only been introduced this year
by \cite{Giorgi:2020ujd} for the Kerr-Newman problem, in a spacetime that was 
discovered over 50 years ago.  Hopefully, results will be quicker in the STU case, 
with our results providing the first hint of encouragement that similar such gargantuan 
efforts as in \cite{Giorgi:2020ujd} may, in the end, be worthwhile for STU metrics too.

In conclusion, we have provided a proof of
mode stability for a massless scalar
field $\psi$ in the background of a wide class of four-dimensional
asymptotically flat rotating and ``electrically
charged'' supergravity black holes, whose energy-momentum tensor satisfies
the dominant energy condition in general relativity.
We handled the difficulties presented by the existence
of ergo-regions in the original metrics by passing to an auxiliary
metric admitting an everywhere lightlike Killing field
and constructing a scalar field $\Psi$ related to $\psi$
by a non-local integral transformation which satisfies  the
massless wave equation with respect
to the auxiliary metric.  We find that 
the associated energy density of $\Psi$ is not only conserved
but is also non-negative.  Our results open the way to establishing 
decay estimates as obtained in
\cite{Dafermos:2014cua}.

\medskip

\noindent{\bf Acknowledgements}

\medskip

We 
acknowledge helpful conversations with
Mihalis Dafermos, Rita Teixeira da Costa and Claude Warnick
during the course of this work.  MC~is supported in part by DOE 
Grant Award de-sc0013528, the Fay R. and Eugene L. Langberg Endowed Chair and
Slovenian Research Agency No. P1-0306.
CNP~is supported in part by DOE grant DE-FG02-13ER42020.  
BFW is supported 
by NSF grant PHY 1607323, 
by the Institut d'Astrophysique de Paris, 
by Astroparticule et Cosmologie  with French state funds 
under grant ANR-14-CE03-0014-01-E-GRAAL, and by the Swedish Research Council under grant 2016-06596, at the Institut Mittag-Leffler in Djursholm, Sweden during the Fall semester of 2019.

\end{document}